\documentclass[aps,prl,reprint,superscriptaddress,floatfix,amsmath,showpacs]{revtex4-1}

\usepackage{graphicx}
\usepackage{amsmath}
\usepackage{textcomp} 

\begin{document}

\title{Detection of Spin Waves in Permalloy using Planar Hall Effect}

\author{Yuri V. Kobljanskyj}
\affiliation{Faculty of Radiophysics, Taras Shevchenko National University of Kyiv, 01601 Kyiv, Ukraine}

\author{Gennadii A. Melkov}
\affiliation{Faculty of Radiophysics, Taras Shevchenko National University of Kyiv, 01601 Kyiv, Ukraine}

\author{Alexander A. Serga}
\email{serga@physik.uni-kl.de}
\affiliation{Fachbereich Physik and Landesforschungszentrum OPTIMAS, Technische Universit\"at Kaiserslautern, 67663 Kaiserslautern, Germany}

\author{Andrei N. Slavin}
\affiliation{Department of Physics, Oakland University, Rochester, Michigan 48309, USA}

\author{Burkard Hillebrands}
\affiliation{Fachbereich Physik and Landesforschungszentrum OPTIMAS, Technische Universit\"at Kaiserslautern, 67663 Kaiserslautern, Germany}

\date{\today}

\begin{abstract}

Rectification of microwave oscillations of magnetization in a permalloy film is realized using planar Hall effect. Two different rectified signals are obtained: a signal from the linearly excited uniform magnetization precession at the frequency of the external pumping and a signal from the pairs of contra-propagating short-wavelength spin waves parametrically generated at a half of the pumping frequency. The second, most unusual, rectified signal is caused by the uniform component of the dynamic magnetization created due to the interference of the phase correlated pairs of parametric spin waves.

\end{abstract}

\maketitle

The experimental investigations of phenomena caused by the spin-orbital interaction in magnetically ordered substances is one of the dominating directions in modern magnetism \cite{Ciccarelli2015}. The spin-orbital effects like anisotropic magnetoresistance (AMR) \cite{McGuire1975}, spin Hall \cite{Hirsch1999} and inverse spin Hall \cite{Saitoh2006} effects allow an experimentalist not only to detect the spin currents caused by linear and nonlinear the microwave magnetization dynamics (see e.g. Ref.~\cite{Kurebayashi2011}), but also to create pure spin currents that are sufficiently large to excite microwave auto-oscillations in magnetic metals \cite{Demidov2012}.
Among the effects that allow one to electrically detect magnetization dynamics in magnetics the \emph{rectification} effects, that are caused by the nonlinear coupling between the spin and charge dynamics and provide resultant signal in the form of a DC voltage, play a particularly important role, since they are sensitive not only to the geometric configuration of the detected microwave fields, but also to the phase and angular relations between the microwave fields and currents \cite{Gui2013}. These effects can be used for the detailed probing of the magnetization dynamics in magnetic micro- and nano-structures \cite{Kubota2007} and, also, for the development of ultra-sensitive microwave detectors \cite{Tulapurkar2005, Bai2008, Prokopenko2013}, demodulation of amplitude-modulated microwave signals \cite{Yamaguchi2007}, and for non-destructive testing \cite{Cao2012}.

Here we present experimental evidence  that another effect caused by the spin-orbital interaction, namely, the \emph{planar Hall effect} (PHE) \cite{Fang2010, Koch1955}, can be successfully used for the detection and rectification of microwave signals exciting the oscillating magnetization dynamics in magnetic metals. In this case, the dependence of the detected DC voltage on the angle between the  microwave magnetic field $\mathbf{h}(t)$, causing the effect, and the bias magnetic field $\mathbf{H}$, magnetizing the magnetic metal, is qualitatively different from the angular dependence of the rectification voltage  caused by the anisotropic magnetoresitance \cite{McGuire1975, Fang2010, Hong1955,Zhu2012}. Moreover, the discovered PHE based rectification process produces DC voltage even in the case when the detected magnetization dynamics is associated with short-wavelength spatially \textit{non-uniform} spin waves parametrically excited in the magnetic metal.

Similarly to the conventional Hall effect, in case of the PHE the flow of the conduction electrons is deflected from the straight propagation path, but this deflection is not caused by the Lorentz force. It occurs due to the interaction of the electron spins with the conductor crystal lattice via the spin-orbital interaction. As a result, the DC electric current $I$ flowing in the plane of  the conductor film creates a voltage drop $V$ between the lateral faces of the film \cite{Koch1955}:
\begin{equation} \label{eq1}
V=I\triangle R(\Theta)=I\triangle R\sin\Theta\cos\Theta \, ,
\end{equation}
where $\Theta$ is the angle between the DC current $\mathbf{I}$ and the magnetization vector $\mathbf{M}$, which is oriented along the bias magnetic field $\mathbf{H}$. In contrast with the conventional Hall effect, the field $\mathbf{H}$ lies in the film plane. The relative value of the PHE related resistance change $\triangle R/R$ has the same order of magnitude as in the AMR case.

\begin{figure}[t]
\includegraphics[width=8.5 cm]{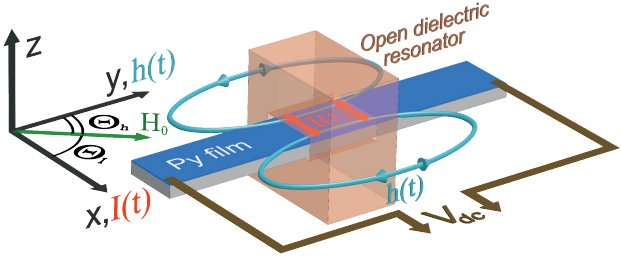}
\caption{\label{fig_setup} (color online). Scheme of the experimental setup for the detection of a rectified  voltage $V_\mathrm{dc}$ caused by the PHE in a permalloy (Py) film sample driven by a microwave pumping. A Py film sputtered on a Si substrate was placed into an axial rectangular opening made in an open dielectric resonator. The DC rectified voltage $V_\mathrm{dc}$ is measured by an oscilloscope. The light blue curves indicate the field lines of the microwave pumping magnetic field $\mathbf{h}$ of the dielectric resonator. The red bold arrows shows the direction of the microwave eddy current $\mathbf{I(t)}$ induced by the pumping. The orientation of the bias magnetic field $\mathbf{H_0}$ in the film plane can be changed relative to the pumping field $\mathbf{h(t)}$ and the eddy current $\mathbf{I(t)}$.}
\end{figure}

Application of a spatially uniform \emph{microwave} magnetic field $\mathbf{h}(t)=\mathbf{h_0}\cos\omega t$ to a magnetic conductor results in two effects.
First, this field excites a spatially uniform microwave eddy current $I(t)$ (see Fig.~\ref{fig_setup}):
\begin{equation} \label{eq2}
I=Ah_0\cos(\omega t - \varphi_I) \,, \quad
\varphi_I=\pi/2 \,,
\end{equation}
where $A$ is a constant determined by the film's material, thickness and boundary conditions.
Second, in the case when the driving frequency $\omega$ is close to the frequency of a ferromagnetic resonance (FMR) in the conductor sample, the magnetization vector $\mathbf{M}$ starts to precess and, thus, the angle $\Theta$ becomes a function of time: \looseness=-1
\begin{equation} \label{eq3}
\Theta=\Theta(t)=\Theta_I+\Theta_p\cos(\omega t-\varphi_R) \,.
\end{equation}
Here $\Theta_I$ is the angle between the eddy current $\mathbf{I}(t)$ and a static component $\mathbf{H_0}$ of the externally applied magnetic field $\mathbf{H}=\mathbf{H_0}+\mathbf{h}$. $\Theta_p$ is the amplitude of the uniform magnetization precession excited by the microwave field. The precession amplitude at the FMR (when $\omega=\gamma\sqrt{H_0(H_0+4\pi M_0)}$) can be found \cite{GurevichMelkov1996} as:
\begin{subequations} \label{eq4}
\begin{eqnarray}
\Theta^{\mathrm{res}}_p & = & \frac{ H_0 + 4\pi M_0}{ H_0 + 2\pi M_0}\frac{h_0}{\triangle H}\sin\Theta_h \,, \\
\varphi_ R& = &\varphi_I  =  \pi/2 \,,
\end{eqnarray}
\end{subequations}
where $\triangle H$ is the FMR linewidth and $M_0$ is the saturation magnetization. $\Theta_h$ is the angle between $\mathbf{h}(t)$ and $\mathbf{H_0}$. It stands that $\Theta_h + \Theta_I = 90^\circ$.

Since in a driven magnetic conductor both the angle $\Theta$ and the current $I$ are functions of time, the voltage drop $V$ in Eq.\,(\ref{eq1}) will get a DC component $V_\mathrm{dc}$. Thus, the incident microwave signal is rectified.
Using Eqs.~(\ref{eq1})--(\ref{eq4}) and assuming a small precession angle $\Theta^{\mathrm{res}}_p \ll 1$, one can obtain the expression for the rectified DC voltage:
\begin{equation} \label{eq5}
V_\mathrm{dc} = \frac{h_0}{2}A\triangle R \Theta_p\cos2\Theta_h .
\end{equation}
It is easy to see from Eq.~(\ref{eq5}) that there is a qualitative difference between the rectification effects caused by the PHE and by the AMR effect: The PHE related DC voltage reaches its maximal magnitude at $\Theta_h=90^\circ$. In contrast, under these conditions the rectified DC voltage caused by the AMR effect vanishes \cite{Fang2010}.

For the experimental investigation of the PHE-related rectification phenomenon we used a specially developed resonator-based technique. The permalloy (Py, $\mathrm{Ni_{81}Fe_{19}}$) film sample was placed into the axial opening made in the center of an open dielectric resonator (see Fig.~\ref{fig_setup}), where the strength of the microwave magnetic field of the $H_{11\delta}$ mode of such a resonator has a maximum \cite{Okaya1962}. The resonator is made of thermostable ceramics with a dielectric permeability $\epsilon \simeq 80$, and has a rectangular shape of the sizes  $3.5\times3.5\times2\,\mathrm{mm}^3$. The opening has the sizes of $1.7\times0.7\times2\,\mathrm{mm}^3$. The resonance frequency of the resonator with the inserted sample is 9350\,MHz. All our measurements were performed at this frequency. In order to excite the resonator mode $H_{11\delta}$ the resonator was placed into the antinodal point of the microwave magnetic field of the waveguide mode $H_{01}$ near the closed end of a conventional rectangular waveguide. The maximum microwave power at the waveguide input reached 100~W. In order to avoid sample heating the experiment was performed in a pulsed regime: microwave pulses of the duration 5~$\mu$s were repeated with a time interval of 20~ms. \looseness=-1

The permalloy film of the sizes of $25\,\mathrm{mm}\times1.5\,\mathrm{mm}\times25\,\mathrm{nm}$ was sputtered on a high-resistance Si substrate of the 0.2\,mm thickness. The film resistance was $100\pm2$\,Ohm. The sample was connected to an oscilloscope for the measurement of the rectified $V_\mathrm{dc}$ voltage (see Fig.~\ref{fig_setup}).

The application of the microwave power $P$ to the waveguide leads to the appearance of on alternating magnetic field $\mathbf{h}$ in the resonator opening.
The preferred orientation of this field in the Py film area is along the $y$ axis (see  Fig.~\ref{fig_setup}) and, consequently, the eddy current $\mathbf{I}$ is directed along the $x$ axis. Due to the PHE, the corresponding electron flow deflects from $\mathbf{x}$ towards the $\mathbf{y}$ direction. This leads to the creation of a rectified voltage $V_\mathrm{dc}$, as it is shown in Fig.~\ref{fig_setup}.

It is worth noting, that the inverse spin Hall effect or the anomalous Hall effect, which could potentially affect the experimental results, in reality do not contribute to the measured voltage $V_\mathrm{dc}$ since, under the given experimental conditions (when the electric current, spins of the conduction electrons, and the induced electric field lie in the same plane), their influence is negligible.

\begin{figure}[t]
\includegraphics[width=8.5 cm]{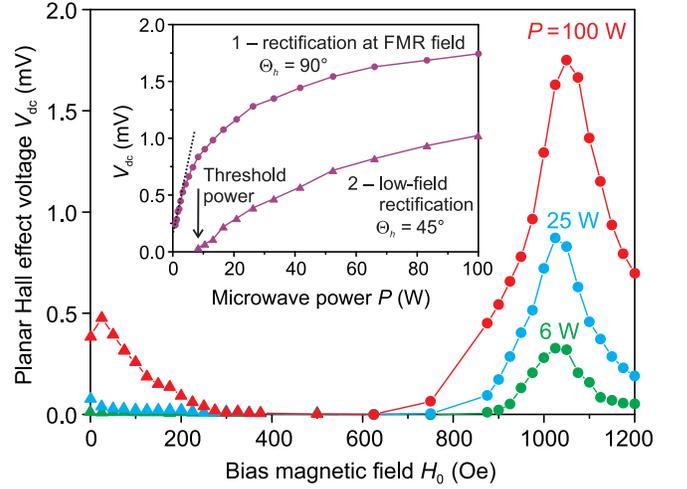}
\caption{\label{fig2} (color online). Dependences of the PHE rectified voltage $V_\mathrm{dc}$ on the magnitude of the bias magnetic field $H_0$ for three input microwave powers: $P = 6$, 25, and 100\,W. $\Theta_h=90^\circ (\mathbf{H_0} \perp \mathbf{h})$. Inset shows dependences of the PHE voltage $V_\mathrm{dc}$ on the input microwave power $P$ in the case of the excitation at the ferromagnetic resonance (curve~1, $H_0=1020$\,Oe) and in the low-field region (curve~2, $H_0=125$\,Oe), respectively.}
\end{figure}

According to Eq.~(\ref{eq5}), the voltage $V_\mathrm{dc}$ is proportional to the amplitude of the uniform precession $\Theta_p$ and, thus, reaches its maximum at the FMR frequency.  Indeed, from Fig.~\ref{fig2}, which represents the experimental dependence of $V_\mathrm{dc}$ on the bias magnetic field $H_0$, one can see the resonance increase of the amplitude of the output rectified voltage $V_\mathrm{dc}$ near the ferromagnetic resonance field $H_0=H_\mathrm{res}=1020$\,Oe.

An interesting and unusual feature of the results presented in Fig.~\ref{fig2} is the appearance of an additional output signal in the region of small magnetic fields $H_0<H_\mathrm{res}$ when the input microwave power $P$ exceeds 8\,W. The nature of this signal will be discussed below.

\begin{figure}[t]
\includegraphics[width=8.5 cm]{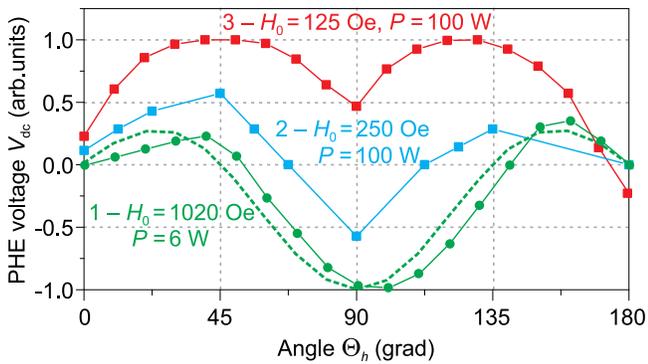}
\caption{\label{fig3} (color online). The green curves show the angular dependence of the PHE voltage $V_\mathrm{dc}$ measured for 6\,W microwave power in the case of excitation at the ferromagnetic resonance ($H_0=1020$\,Oe). Filled green circles -- experiment. The dashed line $\sim\sin\Theta_h\cos2\Theta_h$ is a theoretical curve calculated using Eq.~(\ref{eq5})
The red and blue curves represent the angular dependences measured in the low-field region at$H_0=250$\,Oe (curve~1) and at $H_0=125$\,Oe (curve~2) at the microwave power of $P=100$\,W.
}
\end{figure}

The dependence of the amplitude of the rectified signal $V_\mathrm{dc}$ on the angle $\Theta_h$ between the constant bias magnetic field $\mathbf{H_0}$ and the alternating magnetic field $\mathbf{h}$ in the case of the FMR is shown in Fig.~\ref{fig3}. The experimental curve corresponds well to the angular dependence of the rectified voltage given by Eq.~(\ref{eq5}), which is proportional to $\sim\sin\Theta_h\cos2\Theta_h$.

The absolute value of the FMR-induced output voltage $V_\mathrm{dc}$ measured at $\Theta_h=90^\circ$,  where the FMR uniform precession amplitude $\Theta_p^\mathrm{res}$ reaches its maximum value, is presented in the inset to Fig.~\ref{fig2} (curve~1) as a function of the input microwave power $P$. At low powers $P$ this dependence, shown by the dotted gray line in the inset, is linear. This result follows from Eq.~(\ref{eq5}), where $V_\mathrm{dc} \sim h_0^2 \sim P$. The slope of this linear region on the curve is $2\cdot10^{-4}$\,V/W.
It is worth noting, that this value, which defines the rectification efficiency, is one order of magnitude larger than the previously reported values obtained using the AMR effect \cite{Mosendz2010}. For good agreement with the experiment the constant $A$ in Eq.~(\ref{eq5}) should be chosen to be  0.4\,$\mu$m and, thus corresponds the theoretical estimation $A < 1\mu\mathrm{m}$.

With an increase in the microwave power $P$ above 10~W the rectified voltage saturates. This saturation can be understood as a result of the well-known saturation of the precessional amplitude $\Theta_p$ due to the parametric excitation of short-wavelength spin waves at the frequency of the microwave signal $\omega$ (so-called Suhl's instability of the second order \cite{Suhl1957}).

The inset to Fig.~\ref{fig2} also shows that at low microwave powers $P$ the rectified PHE signal can be observed only around the FMR field $H_\mathrm{res}=1020$\,Oe. With the increase in the input microwave power an additional rectified signal appears in the region of low bias magnetic fields $H_0<250$\,Oe. The angular dependence of the rectified voltage in this low-field region is substantially different from the angular dependence of the signal induced by the FMR. Two of such ``low-field'' angular dependences measured at the maximum microwave power $P=100$\,W are presented in Fig.~\ref{fig3} for $H_0=125$\,Oe and 250\,Oe, respectively.
The maximum of the rectified voltage in this low-field region is observed at $\Theta_h=45^\circ$.
At the same time, no pronounced maximum of the rectified voltage can be observed for the case where the bias and the pumping magnetic fields are perpendicular to each other ($\Theta_h=90^\circ$). In contrast, here, both the magnitude and the polarity of the rectified voltage depend on the magnitude of the bias magnetic field.

\begin{figure}[t]
\includegraphics[width=8.5 cm]{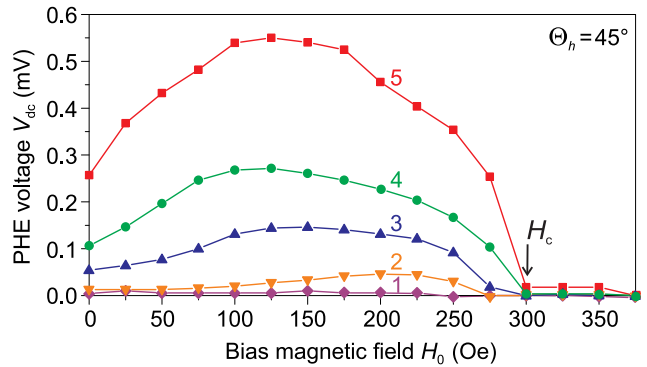}
\caption{\label{fig4} (color online). Magnetic field dependencies of the PHE voltage $V_\mathrm{dc}$ on the microwave power $P$ in the low-field region: 1 -- 6\,W; 2 -- 11\,W; 3 -- 25\,W; 4 -- 50\,W; 5 -- 100\,W. $\Theta_h=45^\circ$.}
\end{figure}

Figure~\ref{fig4} demonstrates the magnetic field dependence of $V_\mathrm{dc}$, which was measured in the low-field area at different microwave powers at the optimum angle $\Theta_h=45^\circ$. It is clear, that the rectified signal appears in a threshold manner, and that the field region of its existence increases with the increasing microwave power. The threshold character of the rectified voltage in the low-field area is also clearly visible from the second curve in the inset in Fig.~\ref{fig2}, which presents the power dependent behavior of the ``low-field'' $V_\mathrm{dc}$ voltage measured at optimum conditions ($\Theta_h=45^\circ \, H_0=125$\,Oe). For microwave powers $P<8$\,W there is no signal. Above this threshold power the function $V_\mathrm{dc}(P)$ increases almost linearly with the slope of  $1.5\cdot10^{-5}$\,V/W, which is approximately one order of magnitude smaller than in the case of the FMR.

The threshold character of the appearance of the ``low-field'' rectified signal can be associated with the parametric excitation of spin waves \emph{at half of the pumping frequency} or, in other words, with the Suhl's instability of the first order \cite{GurevichMelkov1996}. This kind of instability is possible when half of the input microwave frequency lies inside the spin-wave frequency band. In the case of an in-plane magnetized film this condition is satisfied for $H_0<H_\mathrm{c}$,\,where
\begin{equation} \label{eq8}
H_\mathrm{c}=-2\pi M_0 +\left[ (2\pi M_0)^2+\left(\frac{\omega}{2\gamma}\right)^2\right]^{1/2} .
\end{equation}
A calculation shows that under our experimental conditions $H_\mathrm{c}\simeq300$\,Oe. This value corresponds perfectly with the observed value of the "boundary" magnetizing field [see Fig.~\ref{fig4}]. It is also known that the field region of the parametric instability broadens into the direction of smaller bias magnetic fields with the increase in the pumping power. This feature is also visible in Fig.~\ref{fig4}.

The microwave threshold field of the first order parametric instability can be written as  \cite{GurevichMelkov1996}:
\begin{equation} \label{eq9}
h_\mathrm{thr} \simeq \triangle H \frac{\omega}{\omega_M} \frac{1}{\sin2\Theta_k} \,, \quad
\omega_M = 4 \pi M_0
\end{equation}
where $\Theta_k$ is a polar angle of a spin wave (angle between the spin-wave wavevector $\textbf{k}$ and the bias magnetic field $\mathbf H_0$) parametrically excited at half of the microwave pumping frequency $\omega$.
Under our experimental conditions the calculated threshold power $P_\mathrm{thr} \sim h^2_\mathrm{thr}$ is about 10\,W in the optimum situation when $\sin2\Theta_k=1$. This value is close to the experimentally measured threshold power, which is about 8\,W.

Thus, we can conclude that the rectified PHE voltage $V_\mathrm{dc}$ in the low-field area is determined by the parametric excitation of spin waves of the frequency $\omega_k=\omega/2$ and wavenumber $k\geq10^4$\,cm$^{-1}$ lying in the film plane \cite{GurevichMelkov1996}. This conclusion is completely non-trivial, because previously it was assumed that the reason for the appearance of the PHE-based and the AMR-based rectification effects is the uniform magnetization precession ($k=0$) in the film plane (not necessarily along the film thickness \cite{Gui2013}) excited at the frequency of the external pumping field $\omega$. In the case of parametrically excited short spin-waves both these conditions are violated.
As a result, the direct interaction of the uniform eddy current $I$ with the dynamic non-uniform magnetization of short-wavelength spin waves becomes impossible, because their interaction integral vanishes if the integration is done over a macroscopic sample with a size $\gg 2\pi/k$. Moreover, no constant voltage can result from such an interaction due to the fact that the frequency $\omega$ of the eddy current $I$ differs from the frequency $\omega/2$ of the dynamic magnetization of the parametrically excited spin waves.

However, it is well known \cite{Lvov1994} that two parametric spin waves with wave vectors $\mathbf{k}$ and $-\mathbf{k}$ pointing in opposite directions produce an additional dynamic magnetization, which is determined by the product $a_k a_{-k}$ of their amplitudes. This magnetization satisfies all the rectification conditions: it is uniformly distributed in space and oscillates exactly with the pumping frequency $\omega$. The product $a_k a_{-k}$, which is known as an anomalous correlator, can significantly exceed the intensity of the uniform precession $\Theta_c^2 \equiv a_0^2$ \cite{Lvov1994}:
\begin{equation} \label{eq10}
|a_k a_{-k}| = |a_k|^2 \simeq |a_0|^2 (h_0/h_\mathrm{thr})  .
\end{equation}

Moreover, at the moment when the threshold value of $h$ is reached, the phase of the anomalous correlator (and, consequently, the phase $\varphi_R$) becomes equal to $\pi/2$. As a result, the  spin-wave rectification effect appears under the optimum phase condition $\varphi_ R = \varphi_I$. A further increase of the microwave power leads to a shift in the phase of the anomalous correlator from this optimum value \cite{Lvov1994}, and, therefore, results in the saturation of the $V_\mathrm{dc}$ voltage (see curve~2 in the inset in Fig.~\ref{fig2}).

Finally, let us discuss the angular dependences of $V_\mathrm{dc}$ in the low-field area shown in Fig.~\ref{fig3}. In fact, the change of the angle $\Theta_h$ leads to a transition from a parallel ($\Theta_h=0$) to a perpendicular ($\Theta_h=90^\circ$) pumping regime \cite{Neumann2009}. In the case of parallel pumping the alternating magnetization created by the anomalous correlator $a_k a_{-k}$ is directed along the bias field $H_0$ \cite{Lvov1994}. In this case $\triangle R = 0$ \cite{Hong1955, Fang2010} and $V_\mathrm{dc}=0$, which is confirmed in the experiment (see Fig.~\ref{fig3} at $\Theta_h = 0^\circ$ and $180^\circ$). Under the perpendicular pumping the product $a_k a_{-k}$ has a component of the alternating magnetization that is perpendicular to $H_0$ and, thus, is able to contribute to the rectification process. The value of this component is proportional to $\sin2\Theta_k$ \cite{Suhl1957}, where $\Theta_k$ is the polar angle of the parametrically excited spin waves. This angle strongly depends on the magnitude of the bias magnetic field $H_0$ and can be responsible for the field-dependent variation of $V_\mathrm{dc}$, which is visible in Fig.~\ref{fig3} at $\Theta_h=90^\circ$.

In conclusion, the generation of a DC voltage $V_\mathrm{dc}$ caused by the rectification of a microwave signal due to the planar Hall effect has been experimentally observed and measured in a thin ferromagnetic film using a specially developed resonator-based technique. It is found that at low amplitudes of the microwave pumping  the rectified signal is caused by the linearly excited uniform (FMR) magnetization precession, and this signal has a resonance-like dependence on the bias magnetic field. 

When the amplitude of the microwave pumping increases, an additional rectified signal appears in a threshold fashion in the region of low bias magnetic fields. This additional signal is caused by the pairs of parametrically excited short-wavelength spin waves (or magnons) propagating in the film plane with wavenumbers $k\geq10^4$\,cm$^{-1}$ at half of the frequency of the external microwave signal. It should be stressed, that this phenomenon is not associated with the excitation of single magnons, but rather with the excitation of magnon pairs $a_k$ and $a_{-k}$ having a non-zero anomalous correlator $a_k a_{-k}$.

Support by EU-FET InSpin 612759 and by the Ukrainian Fund for Fundamental Research is gratefully acknowledged. This work was also supported in part by the grant from DARPA MTO/MESO grant N66001-11-1-4114, Grant No.ECCS-1305586 from the National Science Foundation of the USA, and by contracts from the US Army TARDEC, RDECOM.

\end{document}